\documentclass[cits]{PoS}
\usepackage{amssymb,fontenc,times,mathptmx,graphicx}
\usepackage{epstopdf}

\title{Renormalization group invariance in the Pinch Technique }

\ShortTitle{Renormalization group invariance }

\author{\speaker{John Cornwall}\\
        University of California at Los Angeles (UCLA)\\
        E-mail: \email{cornwall@physics.ucla.edu}}

\abstract{ We show how to construct, using an elementary extension of the Pinch Technique (PT),   all off-shell Green's functions of a non-Abelian gauge theory so that they are  locally gauge-invariant and renormalization-group invariant (RGI), as the S-matrix is, as well as being process-independent, coupling-constant independent (dimensional transmutation) and satisfying QED-like Ward identities.   These PT-RGI Green's functions are those  of the gauge potentials $gA_{\nu}$, which is RGI itself in the PT.  They differ  from the standard PT Green's functions by a simple multiplicative factor that removes their dependence on the renormalization point $\mu$.      Closed quark and ghost loops do not change this.  We outline how to construct an approximate PT-RGI three-gluon vertex with three physical momentum scales but no $\mu$-dependence, possibly phenomenologically useful. PT-RGI Green's functions obey a hierarchy of Schwinger-Dyson equations (SDEs) that is organized by their momentum dependence in the UV, which can be found exactly from this hierarchy using the PT Ward identities.  One of these Ward identities  yields the gluon propagator as a functional of the vertex, allowing for an SDE truncation that is just the opposite of the often-used gauge technique that expresses the vertex approximately as a functional of the propagator, but which mishandles the UV behavior.     Gluonic spin furnishes complications that would take too long to explain fully here, so we give details of this SDE hierarchy   for an analogous vertex and propagator in a modified form of $\phi^3_6$ theory. The PT-RGI property of all off-shell Green's functions, plus other long-known features of NAGTs coupled to quarks, leads to near-realization of the dreams of S-matrix theorists of the sixties:  An effectively finite theory whose symmetry structure (that of an NAGT) follows from unitarity, and having no fixed singularities in angular momentum.  Effective finiteness means that all Green's functions, including matrix elements of condensates, depend neither on a cutoff nor on a renormalization point $\mu$; in conventional treatments, an off-shell Green's function will depend either on one (unrenormalized, cutoff-dependent) or the other (renormalized, $\mu$-dependent).   }

\FullConference{International Workshop on QCD Greens Functions, Confinement
and Phenomenology\\
                 September 5-9,2011\\
                 ECT Trento, Italy}

\begin{document}

\section{Introduction}

Pinch technique (PT) Green's functions \cite{corn076,binpap,cornbinpap} are not quite physical, because they depend on the renormalization point $\mu$; they are not renormalization-group invariant (RGI). 
 Here we give an almost trivial modification, already studied for the PT gluon propagator \cite{corn138}, that makes  every off-shell PT Green's function not only gauge-invariant but also RGI; we call these PT-RGI Green's functions.

It may sound implausible that there are renormalized off-shell Green's functions that are {\em both } gauge-invariant and RGI, even though the S-matrix itself is. Usually, an unrenormalized Green's function (subscript $U$) is related to a renormalized one (subscript $R$) as:
\begin{equation}
\label{renorm}
Q_U(p_i;g_0;\Lambda_{UV})=Z(\mu /\Lambda_{UV};g_0)Q_R(p_i;g_R(\mu );\Lambda_{NAGT},\mu )
\end{equation}
where $\Lambda_{UV}$ is an ultraviolet cutoff,\footnote{It is somewhat easier to illustrate our results with a cutoff than with dimensional regularization; they are equivalent in our discussion.} $\mu$ is a renormalization scale, $g_0,g_R(\mu )$ are the bare and renormalized couplings, $\Lambda_{NAGT}$ is the finite physical mass scale of the NAGT, and $Z$ is some combination of cutoff-dependent renormalization constants.  It appears that the price of finiteness is the appearance in renormalized quantities of an arbitrary mass $\mu$.  

What we claim is that the renormalized PT proper (1PI) two-, three, and
four-point functions divided by $g_R^2(\mu )$ are equal to their unrenormalized counterparts, and therefore both UV-finite and independent of $\mu$. They are also independent of the gauge coupling $g$.  This is equivalent to using in the PT not the canonical gauge potential $A_{\nu}(x)$ but  the RGI gauge potential $\mathcal{A}_{\nu}\equiv gA_{\nu}(x)$.      Of course, all higher-point skeleton graphs are also independent of $\mu$ since they have no primitive UV divergences. These properties follow from the  Schwinger-Dyson equations (SDEs)  for PT-RGI Green's functions. It is simplest to think of the SDEs in a ghost-free gauge although actual calculations may be simpler in standard gauges with ghosts.  (Of course, PT-RGI Green's functions are the same in any gauge.) Infrared (IR) divergences are removed by dynamical mass generation \cite{corn076,binpap,cornbinpap}.  All of this is what would happen for a {\em finite} field theory with a mass gap, requiring neither a UV cutoff or a renormalization mass scale.  Since all PT Green's functions calculated as we advocate are finite and independent of $\mu$ we argue that a pure NAGT is effectively a finite field theory in $d=4$.  

    The full SDE for the three-gluon vertex, for example, is extremely complicated \cite{binpap08} because of spin, and our discussion will be greatly streamlined, simply pointing out the main points, which will largely be illustrated with a spinless but asymptotically-free model in Sec.~\ref{phi36}. We use SDEs in which all vertices in skeleton graphs are fully-dressed (unlike those of \cite{binpap08} which have one bare vertex). This makes the renormalization of the SDEs quite transparent.  At one loop in perturbation theory the PT three-vertex was found long ago \cite{corn099}, and studied more recently in \cite{brodbing}.  We will easily see how dividing by $g_0^2$ yields a PT-RGI Green's function useful in the UV.  It is another question  to find reasonable and phenomenologically-useful approximations  to the  full SDE solutions that, although approximated, are PT-RGI and useful in the IR as well as the UV; we sketch a treatment of this issue, based on introducing a phenomenological non-running dynamical gluon mass.

  It is much simpler to appreciate the general idea \cite{corn112} in massive $\phi^3_6$, known to be asymptotically-free but with no spin complications.  We use a somewhat modified version of this theory, with a weak coupling to a ``photon" that allows for a QED-like Ward identity, just as for the PT in an NAGT.  The Ward identity is crucial: First, it relates this vertex to the  proper self-energy, so knowing the vertex means knowing the propagator.  Second, because of cancellations, the Ward identity allows us to use free vertices and massive propagators as {\bf input} to the one-loop skeleton graph, yielding an {\bf output} vertex and propagator showing the exact leading UV behavior.  It follows that the output, when re-inserted into the vertex equation, is UV self-consistent, and approximate IR self-consistency comes from the $\phi$-field mass.   This result holds because every skeleton graph with $N>1$ loops is UV non-leading with an asymptotic behavior of the same functional form as implied by the renormalization group at $N$ loops \cite{corn112}.

We believe, although do not prove here, that all these results continue to hold in an NAGT.
Broadly speaking, one modifies  the known PT result for the perturbative three-vertex \cite{corn099} by changing massless propagator denominators to massive ones and adding seagulls.\footnote{Actually,  in an NAGT there are several three-gluon vertices   in the background-field Feynman gauge; the one of \cite{corn099} couples to three background gauge potentials, while that of \cite{binpap08} has only one background potential.}      The output propagator is quite close to an earlier form based on the propagator SDE and the gauge technique\footnote{The gauge technique gives an approximate form for the vertex as a functional of the propagator, valid for small momenta but inaccurate in the UV.  Our present approach derives the propagator from the vertex, and is UV-exact.}.    Our conclusion is that there is a self-consistent realization of the UV behavior of all PT-RGI SDEs that need renormalization.  Since all skeleton graphs beyond the two-, three-, and four-point functions are finite, one further concludes that it is possible to express this UV behavior without ever mentioning a renormalization scale for the renormalized Green's functions.  Both the UV cutoff and the renormalization scale $\mu$ disappear in  all PT-RGI Green's functions.  In this sense, the result is an effectively finite field theory.  

Throughout this paper we work in a Euclidean framework.

\section{The importance of PT-RGI Green's functions}

\subsection{The importance for present-day NAGT applications}

Although the S-matrix is in principle independent  of both $\Lambda_{UV}$ and $\mu$, in standard practice (see, for example, \cite{glover}) the perturbative UV S-matrix   is calculated approximately from Feynman graphs that themselves depend on $\mu$, with cancellation of the $\mu$-dependence that is supposed to become more complete the higher the order of the calculation.  This has led to much confusion about setting scales, leading for example to the idea that one should renormalize in a {\em process-dependent} way.  But it is not easy to learn about a second process from fitting scales to some semi-phenomenological description of a first process.  We suggest that workers in multi-loop QCD processes in the UV do perturbation theory for PT-RGI Green's functions, so that $\mu$ never appears even in an approximation to the S-matrix. This does not mean that the $\mu$-independent PT-RGI algorithms presented here  are any more accurate than standard ones.  

Phenomena of non-perturbative QCD are completely determined by the SDEs' IR properties.  Usually the infinite SDE hierarchy is truncated with the gauge technique that approximately expresses gluon vertices in terms of the gluon propagator; it is accurate in the IR but not in the UV.  We suggest that a better truncation uses the PT-RGI Ward identities to express the propagator in terms of the three-vertex; this is exact at all momenta.  The SDEs are truncated at some finite number of dressed loops.  This is a SDE truncation that is more complex than any yet explored.

\subsection{The importance as part of the realization of the old dreams of S-matrix theory}

Recall (if you are old enough) the maxims of S-matrix theory:
\begin{enumerate}
\item Analyticity   and on-shell unitarity would lead uniquely to the calculation (except for an overall mass scale) of all properties of hadrons, including their masses, couplings, and symmetries.   
\item No infinities would arise in these calculations, as they did in field theory.
\item Nuclear democracy:  No hadron was to be elementary (as they were in field theories of the day); all would be  composites of each other.  This meant, among other things, that all hadrons lay on Regge trajectories, and Born terms somehow vanished.
\end{enumerate}
After a while it became clear that the S-matrix theorists had no way to realize these visions, and (gravity aside) field theory rose once again to dominance.\footnote{Of course, if one quantizes gravity, string theory, which is an S-matrix theory, could once again rise to the top.}  Let us ask now, after decades of work on NAGTs, how close local field theories of spins $\leq 1$ come to realizing the S-matrix vision.  Some parts of the answer come from earlier work, and we add one new part: 
\begin{enumerate}
\item  A series of works \cite{gell,corn025,gris} showed that in a gauge theory any particle with gauge charge lay on a Regge trajectory in perturbation theory.  Long ago, Reggeization and nuclear democracy found expression in an approach called $Z=0$ (for a comprehensive review, see \cite{hayashi}),  which  were supposed to be explained by choosing parameters of a field theory so that the renormalization constants vanished.    It turns out that asymptotic freedom for the PT-RGI gluon propagator is equivalent to $Z=0$, because this propagator, supposed to behave like $Z/p^2$ at large momentum, vanishes more rapidly \cite{corn138}.
 \item The authors of \cite{corn041,lsmith} showed that tree-level unitarity alone was enough to conclude that the only viable theories with spins 0, 1/2, and 1 were Abelian and non-Abelian gauge theories, possibly with a Higgs mechanism.  No particular NAGT could be singled out from unitarity alone.
\item What we add to this list is that PT-RGI Green's functions effectively define NAGTs as finite field theories, because no such Green's function depends on a renormalization scale, and embodies nuclear democracy in that Born terms get absorbed in terms coming from loops in the SDEs.
\end{enumerate}

\section{Schwinger-Dyson equations, RGI and effectively-finite field theory}

To simplify we omit all matter coupled to the NAGT.  In that case, and in a ghost-free gauge, it is always possible to reduce to only one renormalization constant $Z$ for the NAGT PT Green's functions.  The critical point in this equality of renormalization constants is that the PT Ward identities are QED-like.  Note that this renormalization with a {\em single} renormalization constant, under the constraint of gauge invariance, is only possible with the PT.  But it is very important as a matter of principle, if not as a calculational technique, that carrying out the PT procedure in a ghost-free gauge such as the light-cone gauge \cite{corn076} also yields the necessary equality of renormalization constants, because the Ward identities are also QED-like.  In a ghost-free gauge  the Green's functions differ from those in the PT by extra gauge-dependent terms, and it is precisely the task of the PT to remove these terms by recombining the original gauge-dependent Green's functions.  The main tool in this recombination is the ghost-free Ward identities, which are consistent with the PT-RGI property.  Therefore, although the SDEs of the PT have new terms added to those of the ghost-free gauges,   these do not change the renormalization properties needed for PT-RGI Green's functions.

We will  renormalize the two-, three-, and four-point PT functions using a single renormalization mass; for example, for the three-point function we renormalize, via $\Gamma_R(p_i)=Z(\mu )\Gamma_U(p_i)$ with renormalization at the point $p_i^2=\mu^2 \gg \Lambda_{NAGT}^2$.  This way there is only one renormalization constant $Z$ for all three gluonic Green's functions.  The crucial renormalization relations are:
\begin{equation}
\label{renormrel}
d_U=Zd_R;\;\;\Gamma_U=\frac{\Gamma_R}{Z};\;\;\Gamma_U^{(4)}=\frac{\Gamma_R^{(4)}}{Z};\;\; g_0^2=\frac{g_R^2}{Z}.
\end{equation}
Here the subscript $U$ refers to radiatively-corrected unrenormalized Green's functions and $R$ to their renormalized counterparts; $g_0$ is the bare coupling; $d$ is the PT propagator (with a trivial gauge part omitted), $\Gamma$ is the PT three-point vertex, and $\Gamma^{(4)}$ the proper four-point vertex.  We omit vector and group indices, so the same notation serves for the modified $\phi^3_6$ model of Sec.~\ref{phi36} (where $\Gamma^{(4)}$ is unnecessary).   These relations tell us that certain combinations  are RGI-invariant as well as gauge-invariant, as shown in the next equation which also introduces notation for these RGI functions:
\begin{eqnarray}
\label{rginot}
\Delta_R & = &  g_0^2d_U = g_R^2d_R, \\ \nonumber
G_R(p_i) & = & \frac{\Gamma_U(p_i)}{g_0^2}=\frac{\Gamma_R(p_i)}{g_R^2}, \dots
\end{eqnarray}
This is rather remarkable, since the essence of renormalization is that one trades the cutoff $\Lambda_{UV}$ for the renormalization scale $\mu$ [see Eq.~(\ref{renorm})].  We no longer need the subscript $R$.

The three-gluon PT Green's function $\Gamma (p^2_1,p^2_2,p^2_3)$, first constructed in perturbation theory at one loop in  \cite{corn099}, is a good first step toward constructing  a useful ingredient for the analysis of various hadronic processes, as recognized in \cite{brodbing}.  This vertex function is gauge- and process-independent, unlike other proposals in the literature.  It should be possible, as pointed out in \cite{brodbing}, to define three physical scales for any process involving this vertex, instead of trying to find a (generally process-dependent) single scale for $\mu$ that somehow best fits the physics.  

Unfortunately this original PT three-gluon vertex has two drawbacks:  1) it is $\mu$-dependent; 2) it is perturbative and therefore has IR singularities.  In Sec.~\ref{nagtt} we briefly sketch two modifications that should make it directly useful in phenomenological studies.  The first is to divide the vertex by $g^2(\mu )$ so that it is no longer dependent on this renormalization scale; it depends only on three physical scales.  The second, drawing from extensive experience with the PT, is to give the NAGT gluon a dynamical mass.   Satisfaction of a PT Ward identity requires modification of the Green's functions beyond simply adding a mass to the gluon propagator.  The vertex has an extra term containing massless scalar excitations, suggested long ago, that is well-understood \cite{corn090}.  

\section{\label{nagtt} A brief sketch of the NAGT treatment}

\begin{figure}
\begin{center}
\includegraphics[width=5in]{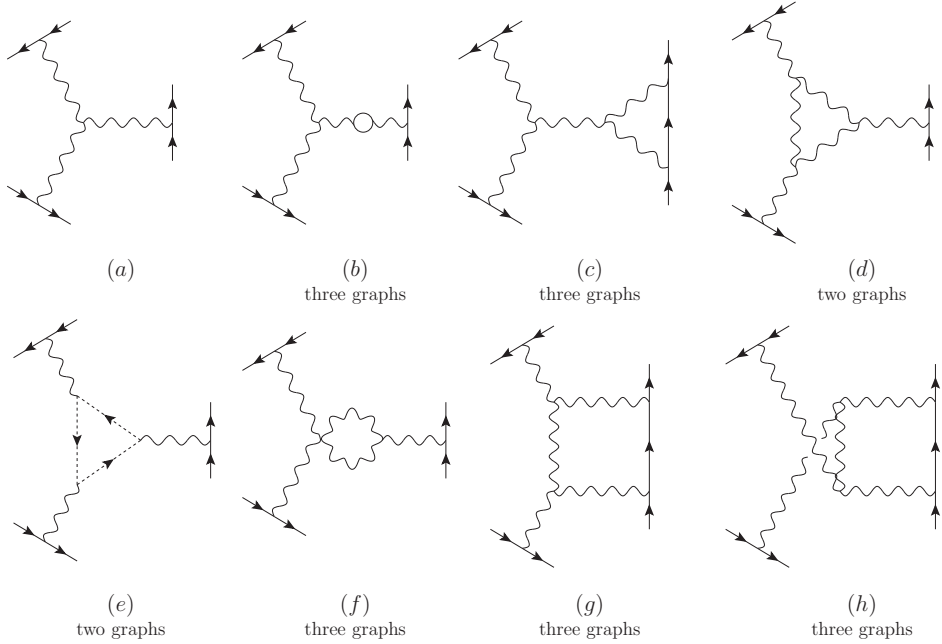}
\caption{\label{3gv} The one-loop perturbative graphs for   the NAGT  PT three-gluon improper vertex, before pinching.   The vertex is extracted from an S-matrix element with three on-shell $q\bar{q}$ pairs. The dotted lines are ghost lines.  Note that only Fig.~(d) has an analog in $\phi^3_6$.[From \cite{cornbinpap}.]}
\end{center}
\end{figure}

The one-loop perturbative S-matrix graphs from which the improper PT three-gluon vertex comes are shown in Fig.~\ref{3gv}.
Not only are there very considerable spin complications, there are many graphs to deal with, even at one loop.  The full expression for this vertex is found in \cite{corn099}, and it was thoroughly analyzed in \cite{brodbing}.  The PT proper vertex comes from stripping off one-loop PT propagators in these graphs.

The proper vertex $G$  satisfies a QED-like Ward identity (group indices suppressed):
\begin{equation}
\label{wident}
p_{1\alpha}G_{\alpha\mu\nu}(p_1,p_2,p_3)=\Delta^{-1}(p_2)P_{\mu\nu}(p_2)-\Delta^{-1}(p_3)P_{\mu\nu}(p_3)
\end{equation}
where $\Delta (p)$ is the scalar coefficient in the PT-RGI propagator of the projector 
$$  P_{\mu\nu}(p)=\delta_{\mu\nu}-\frac{p_{\mu}p_{\nu}}{p^2}. $$
This Ward identity says that scalar  terms in the vertex that contribute have UV asymptotics inverse to   $p^2\Delta (p)$.  An early result for the PT propagator with a dynamical mass \cite{corn076}  $\Delta$ is:
\begin{eqnarray}
\label{corn82}
d_{\mu\nu}(p)  & = & P_{\mu\nu}(p)d (p)+\dots ;\;\;d(p)=\frac{1}{(p^2+m^2)[1+bg^2(\mu )
\ln (\frac{p^2+4m^2}{\mu^2})]}\\ \nonumber
\Delta(p)\equiv g^2(\mu )d(p)& = & \frac{1}{(p^2+m^2)b\ln [\frac{p^2+4m^2}{\Lambda_{NAGT}^2}]}
\end{eqnarray} 
[using Eq.~(\ref{renormch}) below].
The omitted term in Eq.~(\ref{corn82}) is an irrelevant gauge-fixing term that receives no radiative corrections.  The mass should run with momentum, but we ignore that nicety.  This form of $d(p)$ reduces to the one-loop PT propagator at $m=0$, and has the conventional normalization $d^{-1}(p^2=\mu^2)=\mu^2$ for $\mu^2\gg m^2$.  With finite $m$ it is not useful in the timelike regime ($p^2<0$), because it has an unphysical singularity, but this is easily removed \cite{corn138} and we will ignore this nicety too.  The scalar propagator $\Delta$ is, as required, independent of $\mu$, but it is not conventionally normalized to have unit residue at the pole ($p^2=-m^2$), a point that must be remembered when constructing the S-matrix.  The obvious factorization of $\Delta$ is $\Delta (p)=\bar{g}^2(p)H(p)$, where $H(p^2)=1/(p^2+m^2)$ is RGI.  This and Eq.~(\ref{corn82}) define the running charge. Note that, with all IR divergences removed by dynamical mass generation, the running charge is {\em uniquely defined at all momenta} by Eq.~(\ref{corn82}) and the form of $H(p)$.  There is no worrisome scheme dependence \cite{blm}.   
The propagator asymptotic behavior $\sim 1/(p^2\ln p^2)$, shows the often-cited, beginning with \cite{corn076}, failure of positivity for the imaginary part of the propagator, but both $H$ and $\bar{g}^2$ have purely positive imaginary parts; see \cite{corn138}. 

There is one fine point.  Note that $d_{\mu\nu}(p)$ has a pole at $p^2=0$ as long as $m\neq 0$.  This pole, as is well-known \cite{corn076,binpap,cornbinpap}, is a Goldstone-like massless scalar excitation that is necessary for preservation of gauge invariance for massive gauge gluons.  There must be similar poles in the vertex function; the full structure of these was given long ago \cite{corn090}.

The beta-function following from the massive form in Eq.~(\ref{corn82}) is:
\begin{equation}
\label{betafunct}
 \beta (g) = -bg^3 [1-(4m^2/\Lambda_{NAGT}^2)\exp (-1/bg^2)].  
\end{equation}
It approaches the perturbative beta-function as $m$ or $g\rightarrow 0$, and shows conformal behavior ($\beta \approx 0$) near zero momentum, corresponding to $\bar{g}^2(p^2=0)= 1/(b\ln (4m^2/\Lambda_{NAGT}^2)$.   In this limit the renormalized charge is:
\begin{equation}
\label{renormch}
\frac{1}{g^2(\mu )}= b\ln [\mu^2/\Lambda_{NAGT}^2]
\end{equation}
and the corresponding bare charge is:
\begin{equation}
\label{gzero}
\frac{1}{g_0^2}=b\ln (\frac{\Lambda_{UV}^2}{\Lambda_{NAGT}^2})+\dots
\end{equation}
where omitted terms behave like $\ln \ln \Lambda^2_{UV}$ or smaller.

The perturbative RGI-PT vertex \cite{corn099} gives the needed UV behavior to satisfy the Ward identity: 
\begin{equation}
\label{asymbeh}
G\sim b\ln (\frac{p^2}{\Lambda_{NAGT}^2});\;\;\Delta\sim \frac{1}{bp^2\ln (\frac{p^2}{\Lambda_{NAGT}^2})}
\end{equation}

Finally, an exercise like that described in Sec.~\ref{phi36} below gives the SDE for the PT proper self-energy; schematically, it is:
\begin{equation}
\label{propsde}
\Delta^{-1}(p) =\frac{p^2}{g_0^2}+b\int\!G^2\Delta^2+ \int\! G^{(4)}\Delta + \dots
\end{equation}
[Again we use the form of the self-energy SDE in which the product of a bare vertex and $Z$ is replaced by this product as expressed in the vertex SDE, such as Eq.~(\ref{vertexsde}); this self-energy SDE has infinitely many terms, but all of them have the same RGI properties as the term explicitly shown.]   The UV behavior of the explicit term above is consistent with  the asymptotic behavior of Eq.~(\ref{asymbeh}); the $\ln \Lambda^2_{UV}$ term in the integral cancels the explicit logarithm in Eq.~(\ref{gzero}), leaving a finite RGI result for $\Delta$.  

With this background in NAGT, we turn to a spinless example: Modified $\phi^3_6$.

\section{\label{phi36} Mimicking these features in $\phi^3_6$}

Fortunately, the main points are accessible in a much simpler spinless model in six dimensions.  This is based on, although not the same as, $\phi^3_6$, but like this theory the model is asymptotically free.   Some non-perturbative aspects of $\phi^3_6$ and their SDEs were taken up in \cite{corn112}, and will recur here. While this theory, and our model as well, are ultimately doomed because they are unbounded below, no problems arise in our applications, which are mostly restricted to the one-loop skeleton graph. Our general  remarks about the structure of the SDEs are applicable both to NAGTs and to $\phi^3_6$, except when it comes to non-Abelian gauge invariance. Provided that we make a few changes in the factors multiplying certain integrals in the vertex SDEs  of $\phi^3_6$ to mimic the vital role of gauge invariance in the PT Green's functions of an NAGT, the two- and three-point functions of $\phi^3_6$  well illustrate all the ideas that we claim hold for NAGTs.  

To mimic the dynamical gluon mass of an NAGT we add a mass ``by hand'' to the $\phi$ field, which we call the gluon.   (We take the gluon mass as given, and do not need to inquire as to how it is calculated and renormalized.) The UV renormalization-group structure of $\phi^3_6$ is not quite the same as that of an NAGT \cite{corn112}.  However, by making some {\em ad hoc} changes in the factors multiplying the one-loop $\phi^3_6$ SDE integrals we can produce a modified SDE for the the three-point function and a Ward identity for the two-point function that closely resemble those of an NAGT.  In particular, we  modify some constants in $\phi^3_6$ so that, as with an NAGT,  the two-and three-point renormalization constants are the same, so  that standard Green's functions  divided by $g^2$ are RGI.   We study explicitly only the SDE for the three-vertex; the propagator will be constrained formally in terms of the vertex by giving some of the $\phi$ fields electric charge and coupling them to the photon, but its SDE will not be considered.    The Abelian Ward identity then relates  the two-point and three-point functions, with the crucial equality of wave-function and vertex renormalization constants.  Just as for an NAGT one cannot solve the vertex SDE without knowing the propagator; the best one hopes to do is to guess forms for the vertex and propagator that lead to approximate self-consistency.

\begin{figure}
\begin{center}
\includegraphics[width=4in]{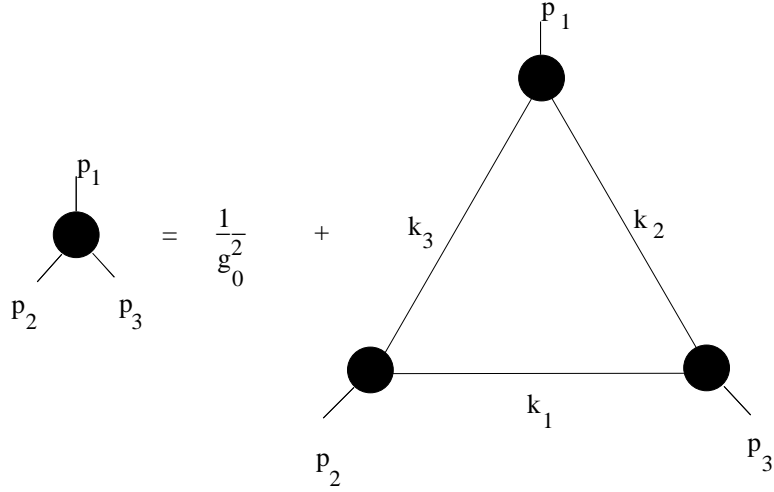}
\caption{\label{sde} The one-loop skeleton graph, with the bare vertex eliminated, for   the $\phi^3_6$ three-gluon proper vertex, divided by $g^2$.  Black circles are the vertex $G$ and lines are the propagator $\Delta$. This graph also appears in the NAGT vertex SDE.}
\end{center}
\end{figure}

The one-loop skeleton graph for the three-gluon vertex equation for  $\phi^3_6$   is shown in Fig.~\ref{sde}. Give these gluons a renormalized mass $m$, which resolves all IR singularities provided that $m>\Lambda /2$, where $\Lambda$ is the theory's mass scale.  (At least for an NAGT it appears that $m\approx 2\Lambda_{NAGT}$.)
With the factor of the integral in the following equation changed  ``by hand'' so that it more closely resembles an NAGT, the model $\phi^3_6$ SDE, or a schema of Fig.~\ref{3gv}(d) in the NAGT SDE, is:
\begin{equation}
\label{vertexsde}
G(p_i)=\frac{1}{g_0^2}-\frac{2b}{\pi^3}\int\!\mathrm{d}^6k\,G(p_1,-k_2,k_3)G(p_2,-k_3,k_1)G(p_3,-k_1,k_2)
\prod_i\Delta(k_i) +\dots
\end{equation}
(all vertex momenta going in). The same general structure holds for the graphs of $d=4$ NAGT (Fig.~\ref{3gv}), where spin numerators yield terms such as $\int\!\mathrm{d}^4k\,k^2$ that are mimicked by the $d=6$ integration.   Think of the factor $b$ as  the one-loop coefficient in the $d=4$ NAGT beta-function $\beta (g)=-bg^3+\dots$, although it is not literally this in $\phi^3_6$.  The infinitely-many omitted vertex skeleton graphs of the model are all functionals of the renormalized Green's functions $G,\Delta$  only and not of $g^2$ or (as we will see) $\mu$. So there is dimensional transmutation, in which the coupling is traded for the physical scale $\Lambda$ of the theory, whether an NAGT or $\phi^3_6$.

In our model we require the same form of the coupling as in Eq.~(\ref{gzero}) for an NAGT except for the labeling of the physical mass scale.  This  amounts to saying that the beta-function is $\beta = -bg^3$ in lowest order.

   We also require that the asymptotic UV behavior for the vertex (when any of its momenta is large and $\mathcal{O}(p)$) and propagator is that given in Eq.~(\ref{asymbeh}) for an NAGT:  {\bf The product $G\Delta$  is the bare propagator  in the UV, to leading order.}   This means  that the vertex and wave-function renormalization constants are the same; this equality will have to be enforced by a Ward identity [see the discussion around Eq.~(\ref{qedward1}) below].       
It also means that the integrand in the SDE behaves as $k^{-6}$ at large $k$, so that there is a UV divergence $\sim \ln \Lambda_{UV}^2$ in the integral. {\bf This divergence is exactly the same as in the perturbative one-loop graph.} Nonetheless, this equation is {\em independent} of $\mu$ because the dependence of the bare coupling $g_0^2$ on $\Lambda_{UV}$ cancels this divergence.   

Since the product $G\Delta$ is $\approx 1/k^2$ for large integration momentum $k$,  both UV self-consistency and IR finiteness come from replacing all products $G\Delta$ by $1/(k_i^2+m^2)$.  Make this replacement to find the approximate vertex:
\begin{equation}
\label{approxvert}
G(p_i)\approx \frac{1}{g_0^2}-b\int\![\mathrm{d}z]\ln [\frac{\Lambda_{UV}^2}{D+m^2}]
\end{equation} 
where 
\begin{eqnarray}
\label{zint}
\int\![\mathrm{d}z] & = & 2\int_0\!\mathrm{d}z_1\,\int_0\!\mathrm{d}z_2\,\int_0\!\mathrm{d}z_3
 \,\delta (1-\sum z_i),\\
D & = & p_1^2\,z_2z_3+p_2^2\,z_3z_1+p_3^2\,z_1z_2.
\end{eqnarray}
(The Feynman parameter $z_i$ goes with the line labeled $k_i$.)  
Combining the last four equations yields an {\em approximation} to the vertex that, in spite of being an approximation, is RGI:
\begin{equation}
\label{approxvert2}
G(p_i)\approx b\int\![\mathrm{d}z]\ln [\frac{D+m^2}{\Lambda^2}].
\end{equation}
Moreover, there is dimensional transmutation because the coupling is absent, and the asymptotic vertex behavior of Eq.~(\ref{asymbeh}) is realized, so that the one-loop skeleton graph is UV consistent with its SDE to leading order in logarithms.  Although we have no space to discuss it here, the approximate vertex is also roughly self-consistent in the IR.  Precisely analogous results hold for the NAGT, where the PT vertex and propagator, divided by $g^2$, are finite, gauge-invariant, and RGI.  

Note that there is no obvious inhomogeneous term in $G$.  Indeed, a recipe for constructing $G$ is to ignore the inhomogeneous term $1/g_0^2$; then replace the UV cutoff by $\Lambda^2$ (assuming, until the calculation is finished, that $\Lambda^2\gg m^2$).  The vertex is as if it were a vertex of a finite theory, needing no renormalization.  The same is true for {\em all} N-point functions.  

Two questions remain:  1)  What about higher orders?  2)  What about the propagator?    The basic argument for higher orders has already been given in   \cite{corn112}.   Every 2PI skeleton graph for the $\phi^3_6$ vertex has girth four (no internal loop of less than four lines), so that no UV logarithms come from Feynman-parameter integrals.  Use Eq.~(\ref{asymbeh}) and standard loop-counting arguments to find that the $N$-loop skeleton graph $G_N$ has the UV behavior:
\begin{equation}
\label{allorduv}
G_N\sim -\int^{\Lambda_{UV}^2}\!\mathrm{d}k^2\frac{1}{k^2 [\ln k^2]^{N-1}},
\end{equation}
which generates precisely the subleading terms in $1/g_0^2$ that accompany $g^{2N+1}$ terms in the beta-function.  [Although not all of the higher-loop skeleton graphs are 2PI, the same UV behavior holds for the vertex SDE in an NAGT.]  Once all these $\Lambda_{UV}$-dependent terms are cancelled, there are new terms in the asymptotic behavior of the vertex that, order by order, are the same (but not necessarily with the same numerical coefficients) as those of an NAGT.  For example, the two-loop graphs contribute a term
$\sim \ln \ln p^2$ to the UV vertex.  

The graphical approach to $\phi^3_6$ reveals a quadratically-divergent propagator.  We bypass this divergence by using  the SDE and Ward identity for the photonic vertex $G_{\alpha}(p_i)$ that holds when we give two of the $\phi$ fields an Abelian ``charge" $\sim g$.  The RGI photonic vertex is defined as $1/g^2$ times the usual proper vertex $\Gamma_{\alpha}$.     The Ward identity once again requires that the UV asymptotics of the leading scalar term of $G_{\alpha}$ be inverse to that of the propagator, so we are again motivated to construct the  one-loop photonic vertex using bare vertices and (massive) propagators, with the coefficient $3b$ of the integral chosen to mimic the NAGT case [see Eq.~(\ref{uvlim}) below]:
\begin{equation}
\label{photvert}
G_{\alpha}(p_i)=\frac{(p_2-p_3)_{\alpha}}{g_0^2}-3b\int\![\mathrm{d}z]\, \ln [\frac{\Lambda_{UV}^2}{D+m^2}]
 [p_2(1-2z_3)-p_3(1-2z_2)]_{\alpha}.
\end{equation}
In particular, choice of the constant $3b$ allows cancellation of cutoff divergences between the integral and the bare vertex.
In the UV limit $p_i^2\approx p^2 \gg m^2,\Lambda^2$, and after cancellation of the $\Lambda_{UV}^2$ logarithms, one finds:
\begin{equation}
\label{uvlim}
G_{\alpha}(p_i)\rightarrow (p_2-p_3)_{\alpha}b\ln [\frac{p^2}{\Lambda^2}].
\end{equation}

The Ward identity:
\begin{equation}
\label{qedward1}
p_{1\alpha}G_{\alpha}(p_i)=\Delta^{-1}(p_3^2)-\Delta^{-1}(p_2^2)
\end{equation}
 follows from the simple relation:
\begin{equation}
\label{qedward}
p_1\cdot [p_2(1-2z_3)-p_3(1-2z_2)]=[\frac{\partial}{\partial z_2}-\frac{\partial}{\partial z_3}]
[D+m^2].
\end{equation}
We can therefore trivially integrate the Ward identity; only the endpoint terms at $z_2,z_3=0$ contribute. Mass terms can be added as constants of integration.   If this is done for an NAGT the extra vertex part derived in \cite{corn090} should be added to the vertex. (The actual mass would, in principle, be determined by combining the vertex and propagator SDEs.) 

It is not necessary to use this approach to the propagator, but it has the advantage of removing what would otherwise be quadratic divergences.  From Eq.~(\ref{qedward})  with no added mass terms:
\begin{equation}
\label{solvewi}
\Delta^{-1}(p_3)=6b\int\!\mathrm{d}z_1\,\mathrm{d}z_2\,\delta (1-z_1-z_2)[D(z_3=0)+m^2]\ln [\frac{D(z_3=0)+m^2}{e\Lambda^2}].
\end{equation}  
After this change, integration and comparison to Eq.~(\ref{qedward1}) yields a finite RGI propagator that looks as if it had come from a finite field theory, with both a kinetic term $\sim p_3^2$  and a mass term.  In the UV limit we find:
\begin{equation}
\label{finalprop}
\Delta^{-1}(p^2)=bp^2\ln [\frac{p^2}{\Lambda^2}],
\end{equation}
the same as the UV limit of the NAGT propagator in Eq.~(\ref{corn82}).

\section{\label{bottom} The bottom line}
\begin{enumerate}
\item All off-shell PT-RGI Green's functions of a matter-free NAGT are gauge-invariant and RGI, behaving as if they were Green's functions of a finite field theory.  This new result, plus older ones, show that NAGTs come very close to realizing the old dreams of S-matrix theory.  These results are best understood in a ghost-free gauge, but will apply to the PT constructed in any gauge.
\item To construct the PT-RGI three-gluon vertex, modify the results of \cite{corn099} by making the gluon propagators massive, according to well-known PT principles \cite{corn076,cornbinpap} that include the addition of certain terms to the vertex containing massless scalars (Goldstone-like excitations) that do not contribute to the S-matrix \cite{corn090}.  The result, although approximate, is gauge-invariant and exactly satisfies the Ward identity, process-independent, RGI (independent of any renormalization scale $\mu$), close to self-consistent with the PT SDE for the three-vertex for all momenta, including the deep IR, and meets the criteria of \cite{brodbing,blm} in having three independent {\em physical} scales for phenomenological use. 
\end{enumerate}


\begin{thebibliography}{99}

\bibitem{corn076}  J.~M.~Cornwall, \emph{Dynamical mass generation in continuum QCD}, Phys.\ Rev.\  D {\bf 26}, 1453 (1982).     
\bibitem{binpap} D.~Binosi, J.~Papavassiliou, \emph{Pinch technique:  Theory  and  applications}, Phys.\ Rept.\  {\bf 479}, 1-152 (2009)
  [arXiv:0909.2536 [hep-ph]]. 
\bibitem{cornbinpap} J.~M.~Cornwall, J.~Papavassiliou, and D.~Binosi, \emph{The pinch technique and applications to non-Abelian gauge theories}, Cambridge University Press, Cambridge 2011.
\bibitem{corn138}  J.~M.~Cornwall,
  \emph{Positivity issues for the pinch-technique gluon propagator and their resolution},
  Phys.\ Rev.\  {\bf D80 } (2009)  096001. 
\bibitem{binpap08}  D.~Binosi, J.~Papavassiliou,  \emph{New Schwinger-Dyson equations for non-Abelian gauge theories}, JHEP {\bf 0811 } (2008)  063;  \emph{Gauge-invariant truncation scheme for the Schwinger-Dyson equations of QCD}, Phys.\ Rev.\  {\bf D77 } (2008)  061702.
\bibitem{corn099}   J.~M.~Cornwall, J.~Papavassiliou,  \emph{Gauge invariant three gluon vertex in QCD},
  Phys.\ Rev.\  {\bf D40 } (1989)  3474.
\bibitem{brodbing} M.~Binger, S.~J.~Brodsky,  \emph{The form-factors of the gauge-invariant three-gluon vertex}, Phys.\ Rev.\  {\bf D74 } (2006)  054016.
\bibitem{corn112} J.~M.~Cornwall, D.~A.~Morris,  \emph{Toy models of nonperturbative asymptotic freedom in $\phi^3$ in six dimensions},
  Phys.\ Rev.\  {\bf D52 } (1995)  6074-6086.
\bibitem{glover}  E.~W.~N.~Glover,  \emph{ Progress in NNLO calculations for scattering processes},
  Nucl.\ Phys.\ Proc.\ Suppl.\  {\bf 116 } (2003)  3-7.
\bibitem{gell}  M.~Gell-Mann, M.~L.~Goldberger, F.~E.~Low {\it et al.},  \emph{Elementary particles of conventional field theory as Regge poles. IV}, Phys.\ Rev.\  {\bf 133 } (1964)  B161-B174.
\bibitem{corn025}  J.~M.~Cornwall, \emph{ Problems of gauge invariance in composite-particle theory},
  Phys.\ Rev.\  {\bf 182 } (1969)  1610-1616.
\bibitem{gris} M.~T.~Grisaru, H.~J.~Schnitzer, H.~-S.~Tsao, \emph {Reggeization of elementary particles in renormalizable gauge theories - vectors and spinors},
  Phys.\ Rev.\  {\bf D8 } (1973)  4498-4509;  J.~M.~Cornwall, G.~Tiktopoulos,  \emph {Infrared behavior of non-Abelian gauge theories}, Phys.\ Rev.\  {\bf D13 } (1976)  3370;  P.~Carruthers, F.~Zachariasen, \emph {Diffraction Scattering in Quantum Chromodynamics},
  in  \emph{ Coral Gables 1976, Proceedings, New Pathways In High-energy Physics, Vol.II}, ed. A. Perlmutter, Plenum, New York 1976, 265-285; V.~S.~Fadin, E.~A.~Kuraev, L.~N.~Lipatov, \emph{ On the Pomeranchuk singularity in asymptotically free theories}, Phys.\ Lett.\  {\bf B60 } (1975)  50-52.
\bibitem{hayashi}  K.~Hayashi {\em et al.}, \emph{Compositeness criteria of particles in quantum field theory and S-matrix theory}, Fortschritte der Physik {\bf 15} (1967) 625.
\bibitem{corn041} J.~M.~Cornwall, D.~N.~Levin, G.~Tiktopoulos, \emph{ Uniqueness of spontaneously broken gauge theories},  
  Phys.\ Rev.\ Lett.\  {\bf 30 } (1973)  1268-1270;   \emph{ Derivation of gauge invariance from high-energy unitarity bounds on the S-matrix}, Phys.\ Rev.\  {\bf D10 } (1974)  1145.=
 \bibitem{lsmith} C.~H.~Llewellyn Smith,  \emph{ High-energy behavior and gauge symmetry},
  Phys.\ Lett.\  {\bf B46 } (1973)  233-236.
\bibitem{corn090}  J.~M.~Cornwall, W.~-S.~Hou,  \emph{Extension of the gauge technique to broken symmetry and finite temperature},  Phys.\ Rev.\  {\bf D34 } (1986)  585.  
\bibitem{blm}  S.~J.~Brodsky, G.~P.~Lepage, P.~B.~Mackenzie,  \emph{On the elimination of scale ambiguities in perturbative quantum chromodynamics},
  Phys.\ Rev.\  {\bf D28 } (1983)  228.


   

\end{thebibliography}
\end{document}